\documentclass[aps,showpacs,floatfix,nofootinbib]{revtex4-2}
\usepackage{amsmath,amssymb,graphicx,bm}

\begin{document}

\title{Shadow Hamiltonians of structure-preserving integrators for Nambu mechanics}


\author{Atsushi Horikoshi}\email{horikosi@tcu.ac.jp}
\affiliation{Department of Natural Sciences, Tokyo City University, 
Tamazutsumi, Setagaya-ku, Tokyo 158-8557, Japan}


\begin{abstract}%
Symplectic integrators are widely implemented numerical integrators for Hamiltonian mechanics, which preserve the Hamiltonian structure (symplecticity)  of the system.
Although the symplectic integrator does not conserve the energy of the system,
it is well known that there exists a conserving modified Hamiltonian, 
called the shadow Hamiltonian.
For the Nambu mechanics, 
which is a kind of generalized Hamiltonian mechanics, 
we can also construct structure-preserving integrators by the same procedure 
used to construct the symplectic integrators.
In the structure-preserving integrator, however,
the existence of shadow Hamiltonians is nontrivial.
This is because the Nambu mechanics is driven by multiple Hamiltonians and
it is nontrivial whether the time evolution by the integrator 
can be cast into the Nambu mechanical time evolution driven by 
multiple shadow Hamiltonians.
In this paper we present a general procedure to calculate the shadow Hamiltonians of 
structure-preserving integrators for Nambu mechanics,
and give an example where the shadow Hamiltonians exist.
This is the first attempt to determine the concrete forms of the shadow Hamiltonians 
for a Nambu mechanical system.
We show that the fundamental identity, which corresponds to the Jacobi identity in Hamiltonian mechanics, plays an important role in calculating the shadow Hamiltonians
using the Baker-Campbell-Hausdorff formula. 
It turns out that the resulting shadow Hamiltonians have indefinite forms 
depending on how the fundamental identities are used.
This is not a technical artifact, because the exact shadow Hamiltonians 
obtained independently have the same indefiniteness.
\end{abstract}


\maketitle
\section{Introduction}
\label{Introduction}
In 1973, Nambu proposed a generalized Hamiltonian mechanics, 
now called the Nambu mechanics \cite{Nambu}.
He extended the phase space spanned by the canonical doublet ($q,p$) 
to one spanned by a multiplet $(x_1, x_2, ..., x_{N})$, 
where $N\ge 3$, and using the Liouville theorem as a guiding principle, 
he generalized the Hamiltonian $H$ 
to multiple Hamiltonians $(H_1, H_2, ..., H_{N-1})$,
the Poisson bracket to the Nambu bracket, and the Hamilton equations 
to the Nambu equations.
It was later found that the Jacobi identity for the Poisson bracket
should be generalized to the fundamental identity for 
the Nambu bracket \cite{SahooValsakumar,Takhtajan}. 
\par
As an example of the Nambu mechanics, Nambu showed that 
the Euler equations for the free rigid body can be derived 
from the Nambu equations \cite{Nambu}.
We can find some interesting applications of the Nambu mechanics,
e.g. in fluid mechanics \cite{NevierBlender,Suzuki},
string/M-theory \cite{HoMatsuo,Yoneya},
and dynamical systems \cite{Unal,FrachebourgKrapivskyBenNaim,Karasozen}.
However, applications have been limited to some particular systems so far, 
because the systems described by the Nambu mechanics should have 
multiple conserved quantities as the Hamiltonians.
Recently we have proposed a novel approach to the Nambu mechanics, 
the hidden Nambu formalism, 
and revealed that the Nambu mechanical structures are hidden 
in any Hamiltonian systems \cite{HorikoshiKawamura}.
For example, a Hamiltonian system of ($q,p$) with a Hamiltonian $H$
can be described by the $N=3$ Nambu system of $(x_1,x_2,x_3)=(q,p,q^2)$
with the Hamiltonians $H_1=H$ and $H_2=x_3-x_1^2$.  
Here the second Hamiltonian $H_2$ works as a constraint between variables.
This formalism allows us to investigate the properties of the Nambu mechanics
in various systems including semiclassical systems
\cite{Horikoshi1,Horikoshi2,ChandreHorikoshi}.
\par
As for numerical integration algorithms for solving ordinary differential equations, 
a number of structure-preserving integrators have been proposed over the decades
\cite{HairerLubichWanner}.
For Hamiltonian systems, the symplectic integrators are well known, which are
numerical integration algorithms that preserve the Hamiltonian structure of the systems.
The symplectic integrator can be constructed by splitting the Hamiltonian $H$
into some pieces and composing the time evolution operators
associated with those pieces.
Therefore, the symplectic integrator does not conserve the energy, i.e.  
the original Hamiltonian $H$.
However, it is also well known that the symplectic integrator
conserves a modified Hamiltonian, called the shadow Hamiltonian $H_{S}$,
which is slightly different from the original Hamiltonian 
\cite{HairerLubichWanner,FrenkelSmit}.
The shadow Hamiltonians are in general calculated  order by order using 
the Baker-Campbell-Hausdorff (BCH) formula and the Jacobi identity.
For the harmonic oscillator, 
some exact expressions of the shadow Hamiltonians have been found by calculations 
without the BCH formula \cite{HardyOkunbor,Kobayashi}.
Since the existence of the shadow Hamiltonian guarantees that 
the error of the original Hamiltonian does not grow secularly, 
symplectic integrators have been used for long-time simulations of various
Hamiltonian systems \cite{HairerLubichWanner,FrenkelSmit,Hernandez}.
Symplectic integrators with various degrees of precision have been proposed, 
and one can choose the integrators appropriate for the purpose 
(for the Python package, see, e.g. Ref. \cite{Chandre}).
\par
On the other hand, for the Nambu systems, we can also construct 
structure-preserving integrators by the splitting and composing procedure 
used to construct symplectic integrators for Hamiltonian systems \cite{Modin}.
Here we refer to those integrators 
not as symplectic integrators, but using the more general name, 
because the Nambu systems do not have symplecticity.
Similarly to the symplectic integrators, the structure-preserving integrator preserves 
the Nambu mechanical structure of the system but does not conserve the original Hamiltonians.
However, in contrast to the symplectic integrators, it is nontrivial whether the shadow Hamiltonians exist in the structure-preserving integrator.
This is because it is nontrivial that the time evolution 
by the structure-preserving integrator can be described by a time evolution operator 
associated with the shadow Hamiltonians.
Indeed, it has been suggested in Ref. \cite{Modin} that 
in the Nambu system corresponding to the Euler equations for the free rigid body,
there are no shadow Hamiltonians of the structure-preserving integrator constructed 
by splitting all the Hamiltonians. 
\par
In this paper we present a general procedure to calculate the shadow Hamiltonians of 
structure-preserving integrators for Nambu mechanics.
As an example where the shadow Hamiltonians exist,
we study a simple $N=3$ Nambu system, 
a harmonic oscillator described by three variables $(x_1,x_2,x_3)=(q,p,q^2)$,
which is given by 
the hidden Nambu formalism \cite{HorikoshiKawamura,Horikoshi1}.
We calculate the shadow Hamiltonians for this model
using the BCH formula and the fundamental identity,
and also we seek to find their exact expressions 
referring to the exact shadow Hamiltonian of the symplectic integrator
\cite{HardyOkunbor,Kobayashi}.

\par
The outline of the paper is as follows.
In Sect. 2 we give a slightly lengthy review of the symplectic integrators
and related issues to emphasize the similarities and differences between 
the symplectic integrators and the structure-preserving integrators presented later.
In Sect. 3 we review the Nambu mechanics and present
a harmonic oscillator system described by three variables $(x_1,x_2,x_3)=(q,p,q^2)$
as an example of $N=3$ Nambu system.
In Sect. 4 we present in a general form 
how to construct the structure-preserving integrators for the Nambu mechanics
and how to calculate the shadow Hamiltonians.
In Sect. 5, using the $N=3$ harmonic oscillator as an example, we show the details 
of the derivations of the shadow Hamiltonians with some numerical results.
Conclusions are given in the last section.

\section{Hamiltonian mechanics}
\subsection{Hamilton equation}
Consider a one degree of freedom Hamiltonian system consisting of a set of canonical variables:
$\bm{x}=~\!^t~\!(x_{1},x_{2})=~\!^t~(q,p)$.
Let $H(x_1,x_2)$ be the Hamiltonian of the system, then Hamilton's equation of motion 
can be written as
\begin{eqnarray}
\frac{df}{dt}=\{f,H\} \label{HamiltonEq}
\end{eqnarray}
for any function $f(x_1,x_2)$. Here
\begin{eqnarray}
\{A,B\} =\frac{\partial(A,B)}{\partial(x_{1},x_{2})}
=\epsilon_{ij}\frac{\partial A}{\partial x_{i}}\frac{\partial B}{\partial x_{j}}\label{Poisson}
\end{eqnarray}
is the Poisson bracket and $\epsilon_{ij}$ is the 2D Levi-Civita symbol.
\par
For a variable transformation $\bm{\mathit{x}}\to\bm{\mathit{y}}$,
the equation of motion written in the new variable $\bm y$ has the same form as in Eq. (\ref{HamiltonEq}) if the transformation satisfies the condition
\begin{eqnarray}
\frac{\partial y_i}{\partial x_k}\epsilon_{kl}\frac{\partial y_j}{\partial x_l}=\epsilon_{ij}.
\label{Symplectic}
\end{eqnarray}
This is called the symplectic condition.
The time evolution along the exact solution of the Hamilton equation (\ref{HamiltonEq})
satisfies this condition.
Under the exact short time evolution 
$\bm{x}(t)\to\bm{x}(t+h)$, where $h$ is a small time step,
the infinitesimal phase space volume is conserved:
$dx_{1}(t+h)dx_{2}(t+h)=dx_{1}(t)dx_{2}(t)$.
This is the Liouville theorem, which follows from the property that 
the velocity field $\dot{\bm{x}}=d\bm{x}/dt$ is divergence-free,
${\rm div}~\dot{\bm{x}}=0$.
The Liouville theorem can also be derived from the symplectic condition (\ref{Symplectic}).
\subsection{Symplectic integrators}
Among various numerical integration schemes for Hamiltonian systems, the symplectic integrators are the ones that preserve symplecticity 
of the system \cite{HairerLubichWanner,FrenkelSmit}.
We define the Liouville operator $X_{H}$ as
\begin{eqnarray}
X_{H}= \epsilon_{ij}\frac{\partial H}{\partial x_{j}}\frac{\partial }{\partial x_{i}},
\end{eqnarray}
and write the Hamilton equation (\ref{HamiltonEq}) as $df/dt=X_{H}f$,
then the short time evolution of $f(t)$ can be formally written as 
\begin{eqnarray}
f(t+h)=e^{hX_{H}}f(t).
\end{eqnarray}
In the case that the Hamiltonian is separable, $H(x_1,x_2)=T(x_2)+V(x_1)$,
the Liouville operator $X_{H}$ can be split into
$X_{H}=X_{T}+X_{V}$, and each short time evolution can be evaluated as
\begin{eqnarray}
e^{h X_{T}}\begin{pmatrix}x_{1}\\ x_{2}\end{pmatrix}\!\!&=~
\begin{pmatrix}x_{1}+h\frac{\partial T}{\partial x_{2}}\\ x_{2}\end{pmatrix},\label{XT}\\
e^{h X_{V}}\begin{pmatrix}x_{1}\\ x_{2}\end{pmatrix}\!\!&=~
\begin{pmatrix}x_{1}\\ x_{2}-h\frac{\partial V}{\partial x_{1}}\end{pmatrix}\label{VT}.
\end{eqnarray}
Since these are exact expressions, each dynamics is symplectic.
\par
Composing exact solutions (\ref{XT})-(\ref{VT}), 
we can define various symplectic integrators for $X_{H}$.
For example, the well-known position Verlet integrator can be constructed by composing
three exact solutions,
\begin{eqnarray}
\Phi^{TVT}_{h}=e^{\frac{h}{2} X_{T}}\circ e^{h X_{V}}\circ e^{\frac{h}{2} X_{T}}. \label{PV1}
\end{eqnarray}
Since symplecticity is preserved in each time evolution, 
this is a symplectic integrator. 
Using the Trotter formula \cite{FrenkelSmit}, the precision of this integrator 
can be evaluated as 
\begin{eqnarray}
e^{hX_{H}}= e^{\frac{h}{2}X_{T}}e^{h X_{V}}
e^{\frac{h}{2} X_{T}}+O(h^{3}).\label{ST1}
\end{eqnarray}
That is, the position Verlet integrator is a second-order symplectic integrator.
This integrator also  has the important property of time-reversal symmetry: $\Phi^{TVT}_{-h}\circ\Phi^{TVT}_{h}=1$.
In the same way, we can construct another well-known 
second-order symplectic integrator, 
the velocity Verlet integrator: 
$\Phi^{VTV}_{h}=e^{\frac{h}{2} X_{V}}\circ e^{h X_{T}}\circ e^{\frac{h}{2} X_{V}}$.

\subsection{Shadow Hamiltonian}
Although the symplectic integrator dose not conserve the energy 
(the original Hamiltonian $H$),
it is known to conserve a quantity close to the original Hamiltonian. 
This means that the product of exponentials
can be represented by
an effective Liouville operator $X_{\rm eff}$
(taking the position Verlet integrator (\ref{PV1}) as an example),
\begin{eqnarray}
e^{hX_{\rm eff}}= e^{\frac{h}{2}X_{T}} e^{hX_{V}} e^{\frac{h}{2} X_{T}},\label{eL1}
\end{eqnarray}
and there exists a modified Hamiltonian, called the ${\it shadow~Hamiltonian}$ $H_{S}$, 
such that $X_{\rm eff}$ can be written as
\begin{eqnarray}
X_{\rm eff}&=&X_{H_{S}}= 
\epsilon_{ij}\frac{\partial H_{S}}{\partial x_{j}}\frac{\partial }{\partial x_{i}}.\label{L1}
\end{eqnarray}
The functional form of $X_{\rm eff}$ can be obtained using 
the BCH formula \cite{HairerLubichWanner}.
For example, applying the second-order BCH formula to Eq. (\ref{eL1}) yields
\begin{eqnarray}
X_{\rm eff}= X_{T} + X_{V} -\frac{h^2}{24}\Big([X_{T},[X_{T},X_{V}]]-2[X_{V},[X_{V},X_{T}]]\Big)
+O(h^{4}),\label{BCH1}
\end{eqnarray}
where $[X,Y]=XY-YX$ is the commutator.
The commutator of two Liouville operators can be expressed in terms of a single 
Liouville operator, 
\begin{eqnarray}
[X_{A},X_{B}]f&=& X_{A}\{f,B\} - X_{B}\{f,A\}= \{\{f,B\},A\}-\{\{f,A\},B\}\nonumber\\
&=&\{f,\{B,A\}\}=X_{\{B,A\}}f.\label{CL1}
\end{eqnarray}
Here we used the Jacobi identity:
$\{\{f,B\},A\}=\{\{f,A\},B\}+\{f,\{B,A\}\}$ or 
$\{\{f,A\},B\}=\{\{f,B\},A\}+\{f,\{A,B\}\}$.
With either identity, we obtain the same result (\ref{CL1}).
Using Eq. (\ref{CL1}) for the BCH formula (\ref{BCH1}) and 
identifying $X_{\rm eff}=X_{H_{S}}$, 
we obtain the shadow Hamiltonian $H_{S}$
for the position Verlet integrator (\ref{PV1}),
\begin{eqnarray}
H_{S}= H -\frac{h^2}{24}\Big(\{\{V,T\},T\}-2\{\{T,V\},V\}\Big)
+O(h^{4}).\label{SHPV}
\end{eqnarray}
Although the original Hamiltonian $H$ is not conserved in the time evolution by 
the position Verlet integrator $\Phi^{TVT}_{h}$, 
the shadow Hamiltonian $H_{S}$ is conserved.
\subsection{Example: Harmonic oscillator}
We show an example of a harmonic oscillator,
\begin{eqnarray}
H(x_1,x_2)&=&\frac{1}{2m}x_2^{2}+\frac{m\omega^2}{2}x_1^{2}.\label{HO1}
\end{eqnarray}
This Hamiltonian is separable, $H(x_1,x_2)=T(x_2)+V(x_1)$, 
where $T(x_2)=(1/2m)x_2^{2}$ and $V(x_1)=(m\omega^2/2)x_1^{2}$, 
and the Liouville operator $X_{H}$ can be split into
$X_{H}=X_{T}+X_{V}$, where
\begin{eqnarray}
X_{T}= \frac{1}{m}x_{2}\frac{\partial }{\partial x_{1}},~~~~~~
X_{V}= -m\omega^{2}x_{1}\frac{\partial }{\partial x_{2}}\label{XTXV}.
\end{eqnarray}
Here we consider the position Verlet integrator (\ref{PV1}).
Using the BCH formula (\ref{BCH1}), the shadow Hamiltonian (\ref{SHPV}) 
can be evaluated as
\begin{eqnarray}
H_{S}= H -\frac{\omega^2}{24m}h^2x_2^2+\frac{m\omega^4}{12}h^2x_1^2
+O(h^{4}).\label{SHPV1}
\end{eqnarray}
On the other hand, it can be shown by explicit calculation 
that the position Verlet integrator conserves the following quantity \cite{Kobayashi}:
\begin{eqnarray}
H_{c}=H-\frac{\omega^2}{8m}h^2x_2^{2}.
\label{CQPV}
\end{eqnarray}
Although $H_{c}$ is conserved, it is NOT the shadow Hamiltonian.
That is, the Liouville operator associated with $H_c$,
\begin{eqnarray}
X_{H_c}= \frac{1}{m}\Big(1-\frac{\omega^2 h^2}{4}\Big)x_{2}\frac{\partial }{\partial x_{1}}
-m\omega^{2}x_{1}\frac{\partial }{\partial x_{2}},\label{XHc}
\end{eqnarray}
is not the effective operator $X_{\rm eff}$.
The exact shadow Hamiltonian $H_{S}^{e}$ in the  case of the harmonic oscillator
has been derived by some authors \cite{HardyOkunbor,Kobayashi},
\begin{eqnarray}
X_{\rm eff}&=&X_{H_{S}^{e}}=F(\omega h)X_{H_{c}},\label{Xeff1}\\
H_{S}^{e}&=&F(\omega h)H_{c},\label{SHPV2}
\end{eqnarray}
where the factor $F(x)$ is given by
\begin{eqnarray}
F(x)&=&\sum_{n=0}^{\infty}\frac{(n!)^2}{(2n+1)!}x^{2n}~=~
\frac{2\arcsin(\frac{x}{2})}{x\sqrt{1-\frac{x^2}{4}}},\label{factor1}\\
&=&1+\frac{1}{6}x^2+\frac{1}{30}x^4+O(x^{6}),\label{factor2}
\end{eqnarray}
for $0<|x|<2$.
\section{Nambu mechanics}
\subsection{Nambu equation} 
Nambu generalized the Hamiltonian mechanics 
by generalizing the canonical doublet $q$ and $p$ 
to an $N$-plet, where $N\ge 3$ \cite{Nambu}. 
Here we consider an $N=3$ system:
$\bm{x}=~\!^t~\!(x_{1},x_{2},x_{3})$.
Nambu then generalized the Hamilton's equation of motion (\ref{HamiltonEq})
to the Nambu equation,
\begin{eqnarray}
\frac{df}{dt} =\{f,H,G\}.\label{NambuEq}
\end{eqnarray}
Here $H$ and $G$ are Hamiltonians,
and $\{*,*,*\}$ is the Nambu bracket,
\begin{eqnarray}
\{A, B, C\} = 
\frac{\partial (A, B, C)}{\partial (x_1, x_2, x_3)}
= \epsilon_{ijk}
\frac{\partial A}{\partial x_{i}}\frac{\partial B}{\partial x_{j}}\frac{\partial C}{\partial x_{k}},
\label{NB}
\end{eqnarray}
where $\epsilon_{ijk}$ is  the 3D Levi-Civita symbol.
Nambu mechanics is an $N$-variable dynamics driven by $N-1$ Hamiltonians, 
and those Hamiltonians are conserved.
\par
For a variable transformation $\bm{\mathit{x}}\to\bm{\mathit{y}}$,
the equation of motion written in the new variable $\bm y$ has the same form as in Eq. (\ref{NambuEq}) if the transformation satisfies the condition
\begin{eqnarray}
\epsilon_{lmn}\frac{\partial y_i}{\partial x_l}\frac{\partial y_j}{\partial x_m}
\frac{\partial y_k}{\partial x_n}=\epsilon_{ijk}.
\label{Condition1}
\end{eqnarray}
This is similar to the symplectic condition (\ref{Symplectic}), but not the same as it.
The time evolution along the exact solution of the Nambu equation (\ref{NambuEq})
satisfies this condition (\ref{Condition1}).
As with the Hamiltonian mechanics, the Nambu mechanics is divergence-free,
${\rm div}~\dot{\bm{x}}=0$, 
and therefore the Liouville theorem also holds in the Nambu mechanics. 
 

\subsection{Example: $N=3$ harmonic oscillator}
The systems described by the Nambu mechanics have been known 
only in limited examples, 
such as the Euler equations for the free rigid body \cite{Nambu}.
In 2013 we proposed the hidden Nambu formalism and showed that systems 
extended to include composite variables can be described by the Nambu mechanics \cite{HorikoshiKawamura,Horikoshi1}.
As an example, here we consider a harmonic oscillator (\ref{HO1}) 
with $N=3$ variables, one of them being a composite variable,
\begin{eqnarray}
x_1=q,~~~ x_2=p,~~~ x_3=q^{2}.
\label{Ex1}
\end{eqnarray}
Defining two Hamiltonians as
\begin{eqnarray}
H(x_1,x_2,x_3)&=&\frac{1}{2m}x_2^{2}+\frac{m\omega^2}{2}x_3,\label{H}\\
G(x_1,x_2,x_3)&=&x_3-x_1^2,\label{G}
\end{eqnarray}
the Nambu equations  (\ref{NambuEq}) read,
\begin{eqnarray}
\frac{d}{dt}x_{1}=\frac{1}{m}x_{2},~~~~~~
\frac{d}{dt}x_{2}=-m\omega^{2}x_{1},~~~~~~
\frac{d}{dt}x_{3}=\frac{2}{m}x_{1}x_{2}.
\label{NambuOHeqs}
\end{eqnarray}
These are consistent with the Hamilton equations for $q$, $p$, and $q^2$.
The second Hamiltonian $G$ works as a constraint between variables,
$G=x_3-x_1^2={\rm Const}$.
If we take the initial condition $x_3(0)=x_1^2(0)$ satisfying the relation (\ref{Ex1}), 
then the constraint becomes $G=0$.

\section{Numerical integrators for Nambu mechanics}
\subsection{Structure-preserving integrators}
In the Nambu mechanics, the condition in Eq. (\ref{Condition1}) holds 
instead of the symplectic condition (\ref{Symplectic}).
In this paper we refer to the integrators that satisfy the condition  in Eq. (\ref{Condition1})
as structure-preserving integrators.
To see how to construct the structure-preserving integrators,
we consider here an $N=3$ Nambu mechanical system. 
The extension to the $N\ge 4$ system is straightforward.
We define the generalized Liouville operator $X_{H,G}$ as
\begin{eqnarray}
X_{H,G}= \epsilon_{ijk}
\frac{\partial H}{\partial x_{j}}\frac{\partial G}{\partial x_{k}}\frac{\partial }{\partial x_{i}}.
\label{gLiouville}
\end{eqnarray}
Then the Nambu equation  (\ref{NambuEq}) can be written as
$df/dt=X_{H,G}f$
and the short time evolution of $f(t)$ can be formally expressed as 
\begin{eqnarray}
f(t+h)=e^{hX_{H,G}}f(t).
\end{eqnarray}
We assume that two Hamiltonians $H$ and $G$ are separable,
\footnote{In Ref. \cite{Modin} Modin has 
considered an $N=3$ Nambu system with separable Hamiltonians
and presented three types of integrators:
one obtained by splitting $H$, 
one by splitting $G$, and 
one by splitting $H$ and $G$.
In the present paper, we consider only the integrators obtained by splitting all Hamiltonians
$H$ and $G$.
If we split only one Hamiltonian (e.g. $H$) and not the other one ($G$), the resulting 
integrator is equivalent to the integrator for a noncanonical Hamiltonian system
$\frac{df}{dt}=J_{ij}\frac{\partial f}{\partial x_i}\frac{\partial H}{\partial x_j}$, where
the Poisson matrix $J$ is defined by using $G$: 
$J_{ij}=\epsilon_{ijk}\frac{\partial G}{\partial x_k}$.
For the noncanonical Hamiltonian system, the shadow Hamiltonian can be 
calculated using the same procedure as for Hamiltonian systems.
} 
\begin{eqnarray}
H(x_1,x_2,x_3)&=&H_1(x_1)+H_2(x_2)+H_3(x_3),\\
G(x_1,x_2,x_3)&=&G_1(x_1)+G_2(x_2)+G_3(x_3).
\end{eqnarray}
Since $X_{H_1,G_1}=X_{H_2,G_2}=X_{H_3,G_3}=0$,
the Liouville operator $X_{H,G}$ can be split into
\begin{eqnarray}
X_{H,G}=X_{H_1,G_2}+X_{H_1,G_3}+X_{H_2,G_1}+X_{H_2,G_3}+X_{H_3,G_1}+X_{H_3,G_2}.\label{HGs}
\end{eqnarray}
We define the following three Liouville operators
such that $X_{H,G}=X_{1}+X_{2}+X_{3}$,
\begin{eqnarray}
X_{1}&=&X_{H_2,G_3}+X_{H_3,G_2}
=\left(\frac{\partial H_2}{\partial x_2}\frac{\partial G_3}{\partial x_3}
-\frac{\partial H_3}{\partial x_3}\frac{\partial G_2}{\partial x_2}\right)
\frac{\partial }{\partial x_{1}},\label{X1}\\
X_{2}&=&X_{H_3,G_1}+X_{H_1,G_3}
=\left(\frac{\partial H_3}{\partial x_3}\frac{\partial G_1}{\partial x_1}
-\frac{\partial H_1}{\partial x_1}\frac{\partial G_3}{\partial x_3}\right)
\frac{\partial }{\partial x_{2}},\label{X2}\\
X_{3}&=&X_{H_1,G_2}+X_{H_2,G_1}
=\left(\frac{\partial H_1}{\partial x_1}\frac{\partial G_2}{\partial x_2}
-\frac{\partial H_2}{\partial x_2}\frac{\partial G_1}{\partial x_1}\right)
\frac{\partial }{\partial x_{3}}.\label{X3}
\end{eqnarray}
Each Liouville operator gives the exact short time evolution as follows:
\begin{eqnarray}
e^{h X_{1}}\begin{pmatrix}x_{1}\\ x_{2}\\ x_{3}\end{pmatrix}\!\!&=&
\begin{pmatrix}
x_{1}+h \left(\frac{\partial H_2}{\partial x_2}\frac{\partial G_3}{\partial x_3}
-\frac{\partial H_3}{\partial x_3}\frac{\partial G_2}{\partial x_2}\right)\\ x_{2}\\ x_{3}
\end{pmatrix},\label{hX1g}\\
e^{h X_{2}}\begin{pmatrix}x_{1}\\ x_{2}\\ x_{3}\end{pmatrix}\!\!&=&
\begin{pmatrix}
x_{1}\\ x_{2}+h\left(\frac{\partial H_3}{\partial x_3}\frac{\partial G_1}{\partial x_1}
-\frac{\partial H_1}{\partial x_1}\frac{\partial G_3}{\partial x_3}\right)\\ x_{3}
\end{pmatrix},\label{hX2g}\\
e^{h X_{3}}\begin{pmatrix}x_{1}\\ x_{2}\\ x_{3}\end{pmatrix}\!\!&=&
\begin{pmatrix}
x_{1}\\ x_{2}\\ x_{3}+h\left(\frac{\partial H_1}{\partial x_1}\frac{\partial G_2}{\partial x_2}
-\frac{\partial H_2}{\partial x_2}\frac{\partial G_1}{\partial x_1}\right)
\end{pmatrix}.\label{hX3g}
\end{eqnarray}
In each time evolution,
the condition  (\ref{Condition1}) is satisfied and the Liouville theorem holds. 
\par
Composing exact solutions (\ref{hX1g})-(\ref{hX3g}), 
we can define various integrators for $X_{H,G}$.
For example, by composing five exact solutions symmetrically, 
we can construct an integrator,
\begin{eqnarray}
\Phi^{12321}_{h}=e^{\frac{h}{2} X_{1}}\circ e^{\frac{h}{2} X_{2}}
\circ e^{h X_{3}}
\circ e^{\frac{h}{2} X_{2}}\circ e^{\frac{h}{2} X_{1}}. 
\label{12321}
\end{eqnarray}
Since the condition  (\ref{Condition1}) is satisfied by each step, 
the integrator $\Phi^{12321}_{h}$ keeps the condition  (\ref{Condition1}) .
Therefore, $\Phi^{12321}_{h}$ is a structure-preserving integrator. 
This is a second-order integrator because its precision can be evaluated 
using the Trotter formula \cite{FrenkelSmit}, 
\begin{eqnarray}
e^{hX_{H,G}}= 
e^{\frac{h}{2}X_{1}}e^{\frac{h}{2}X_{2}}e^{h X_{3}}e^{\frac{h}{2} X_{2}}e^{\frac{h}{2}X_{1}}
+O(h^{3}).\label{ST2}
\end{eqnarray}
This integrator also has the time-reversal symmetry:
$\Phi^{12321}_{-h}\circ\Phi^{12321}_{h}=1$.
We can construct a total of $3!=6$ types of 
second-order structure-preserving integrators
by composing five exact solutions symmetrically.

\subsection{Shadow Hamiltonians}
The product of exponentials can be represented by
an effective Liouville operator $X_{\rm eff}$. For the integrator $\Phi^{12321}_{h}$,
\begin{eqnarray}
e^{hX_{\rm eff}}= 
e^{\frac{h}{2}X_{1}}e^{\frac{h}{2}X_{2}}e^{h X_{3}}e^{\frac{h}{2} X_{2}}e^{\frac{h}{2}X_{1}}.\label{eL2}
\end{eqnarray}
The shadow Hamiltonians $H_{S}$ and $G_{S}$ are defined as
\begin{eqnarray}
X_{\rm eff}&=&X_{H_{S},G_{S}}= \epsilon_{ijk}\frac{\partial H_{S}}{\partial x_{j}}
\frac{\partial G_{S}}{\partial x_{k}}\frac{\partial }{\partial x_{i}}.
\label{L2}
\end{eqnarray}
To derive the shadow Hamiltonians, let us write $X_{\rm eff}$ 
in the form of Eq. (\ref{L2}).
First of all, 
using the BCH formula \cite{HairerLubichWanner}, $X_{\rm eff}$ can be written as
\begin{eqnarray}
X_{\rm eff}= X_{H,G} \!\!\!&-&\!\!\!\frac{h^2}{24}\Big(~
[X_{1},[X_{1},X_{2}]]+[X_{1},[X_{1},X_{3}]]+[X_{1},[X_{2},X_{3}]]\nonumber\\
&&~-2[X_{2},[X_{2},X_{1}]]+[X_{2},[X_{2},X_{3}]]+[X_{2},[X_{1},X_{3}]]\nonumber\\
&&~-2[X_{3},[X_{3},X_{1}]]-2[X_{3},[X_{3},X_{2}]]+3[X_{3},[X_{1},X_{2}]]
~\Big)+O(h^{4}).\label{BCH2}
\end{eqnarray}
Replacing $X_1$, $X_2$, and $X_3$ by the sum of the original 
Liouville operators (\ref{X1})-(\ref{X3}) 
and using the notation $X_{ij}=X_{H_{i},G_{j}}$,
Eq. (\ref{BCH2}) can be expressed as
\begin{eqnarray}
&& \!\!\!\!\!\!\!\!\!\!\!\!\!\!\!\!\!\!\!\!\!\!\!\!\!\!\!
X_{\rm eff} = X_{H,G}-\frac{h^2}{24}\Big(~~
\nonumber\\
&&[X_{23},[X_{23},X_{31}]]+[X_{23},[X_{23},X_{13}]]
+[X_{23},[X_{32},X_{31}]]+[X_{23},[X_{32},X_{13}]]
\nonumber\\
&&\!\!\!\!\!
+[X_{32},[X_{23},X_{31}]]+[X_{32},[X_{23},X_{13}]]
+[X_{32},[X_{32},X_{31}]]+[X_{32},[X_{32},X_{13}]]
\nonumber\\
&&\!\!\!\!\!
+[X_{23},[X_{23},X_{12}]]+[X_{23},[X_{23},X_{21}]]
+[X_{23},[X_{32},X_{12}]]+[X_{23},[X_{32},X_{21}]]
\nonumber\\
&&\!\!\!\!\!
+[X_{32},[X_{23},X_{12}]]+[X_{32},[X_{23},X_{21}]]
+[X_{32},[X_{32},X_{12}]]+[X_{32},[X_{32},X_{21}]]
\nonumber\\
&&\!\!\!\!\!
+[X_{23},[X_{31},X_{12}]]+[X_{23},[X_{31},X_{21}]]
+[X_{23},[X_{13},X_{12}]]+[X_{23},[X_{13},X_{21}]]
\nonumber\\
&&\!\!\!\!\!
+[X_{32},[X_{31},X_{12}]]+[X_{32},[X_{31},X_{21}]]
+[X_{32},[X_{13},X_{12}]]+[X_{32},[X_{13},X_{21}]]
\nonumber\\
&&\!\!\!\!\!\!\!\!
-2[X_{31},[X_{31},X_{23}]]-2[X_{31},[X_{31},X_{32}]]
-2[X_{31},[X_{13},X_{23}]]-2[X_{31},[X_{13},X_{32}]]
\nonumber\\
&&\!\!\!\!\!\!\!\!
-2[X_{13},[X_{31},X_{23}]]-2[X_{13},[X_{31},X_{32}]]
-2[X_{13},[X_{13},X_{23}]]-2[X_{13},[X_{13},X_{32}]]
\nonumber\\
&&\!\!\!\!\!
+[X_{31},[X_{31},X_{12}]]+[X_{31},[X_{31},X_{21}]]
+[X_{31},[X_{13},X_{12}]]+[X_{31},[X_{13},X_{21}]]
\nonumber\\
&&\!\!\!\!\!
+[X_{13},[X_{31},X_{12}]]+[X_{13},[X_{31},X_{21}]]
+[X_{13},[X_{13},X_{12}]]+[X_{13},[X_{13},X_{21}]]
\nonumber\\
&&\!\!\!\!\!
+[X_{31},[X_{23},X_{12}]]+[X_{31},[X_{23},X_{21}]]
+[X_{31},[X_{32},X_{12}]]+[X_{31},[X_{32},X_{21}]]
\nonumber\\
&&\!\!\!\!\!
+[X_{13},[X_{23},X_{12}]]+[X_{13},[X_{23},X_{21}]]
+[X_{13},[X_{32},X_{12}]]+[X_{13},[X_{32},X_{21}]]
\nonumber\\
&&\!\!\!\!\!\!\!\!
-2[X_{12},[X_{12},X_{23}]]-2[X_{12},[X_{12},X_{32}]]
-2[X_{12},[X_{21},X_{23}]]-2[X_{12},[X_{21},X_{32}]]
\nonumber\\
&&\!\!\!\!\!\!\!\!
-2[X_{21},[X_{12},X_{23}]]-2[X_{21},[X_{12},X_{32}]]
-2[X_{21},[X_{21},X_{23}]]-2[X_{21},[X_{21},X_{32}]]
\nonumber\\
&&\!\!\!\!\!\!\!\!
-2[X_{12},[X_{12},X_{31}]]-2[X_{12},[X_{12},X_{13}]]
-2[X_{12},[X_{21},X_{31}]]-2[X_{12},[X_{21},X_{13}]]
\nonumber\\
&&\!\!\!\!\!\!\!\!
-2[X_{21},[X_{12},X_{31}]]-2[X_{21},[X_{12},X_{13}]]
-2[X_{21},[X_{21},X_{31}]]-2[X_{21},[X_{21},X_{13}]]
\nonumber\\
&&\!\!\!\!\!\!\!\!
+3[X_{12},[X_{23},X_{31}]]+3[X_{12},[X_{23},X_{13}]]
+3[X_{12},[X_{32},X_{31}]]+3[X_{12},[X_{32},X_{13}]]
\nonumber\\
&&\!\!\!\!\!\!\!\!
+3[X_{21},[X_{23},X_{31}]]+3[X_{21},[X_{23},X_{13}]]
+3[X_{21},[X_{32},X_{31}]]+3[X_{21},[X_{32},X_{13}]]
\nonumber\\
&&~~\Big)+O(h^{4}).\label{BCH3}
\end{eqnarray}
In order for $X_{\rm eff}$ to be written in the form of Eq. (\ref{L2}), 
the second-order correction terms in Eq. (\ref{BCH3}) must be represented 
by a sum of Liouville operators. 
In the case of symplectic integrators for the Hamiltonian mechanics,
we used the Jacobi identity to derive Eq. (\ref{CL1}),
whereas in the case of the structure preserving integrators for Nambu mechanics,
 we use the fundamental identity 
\cite{SahooValsakumar,Takhtajan}.
Let us consider to represent $[X_{A,B},[X_{C,D},X_{E,F}]]$
by a sum of Liouville operators, where $A\sim F$ are any Hamiltonians.
First, the commutator of $X_{C,D}$ and $X_{E,F}$ can be written as
\begin{eqnarray}
[X_{C,D},X_{E,F}]f&=& X_{C,D}\{f,E,F\} - X_{E,F}\{f,C,D\}\nonumber\\
&=&\{\{f,E,F\},C,D\}-\{\{f,C,D\},E,F\}.\label{CL2}
\end{eqnarray}
If we use the fundamental identity for the first term in Eq. (\ref{CL2}),
$\{\{f,E,F\},C,D\}=\{\{f,C,D\},E,F\}+\{f,\{E,C,D\},F\}+\{f,E,\{F,C,D\}\}$,
then we obtain 
\begin{eqnarray}
\!\!\!\!\!\![X_{C,D},X_{E,F}]f=\Big( X_{\{E,C,D\},F} + X_{E,\{F,C,D\}}\Big)f.\label{CL2a}
\end{eqnarray}
On the other hand, 
if we use the fundamental identity for the second term in Eq. (\ref{CL2}),
$\{\{f,C,D\},E,F\}=\{\{f,E,F\},C,D\}+\{f,\{C,E,F\},D\}+\{f,C,\{D,E,F\}\}$,
then we obtain 
\begin{eqnarray}
[X_{C,D},X_{E,F}]f=\Big(-X_{\{C,E,F\},D} - X_{C,\{D,E,F\}}\Big)f.\label{CL2b}
\end{eqnarray}
That is, the expression of $[X_{C,D},X_{E,F}]$ is not unique and depends on 
which term in Eq. (\ref{CL2}) the fundamental identity is used for.
Here we choose the first term in Eq. (\ref{CL2}) and proceed with
the expression (\ref{CL2a}). Then we have
\begin{eqnarray}
[X_{A,B},[X_{C,D},X_{E,F}]]f &=&
[X_{A,B},X_{\{E,C,D\},F}]f+[X_{A,B},X_{E,\{F,C,D\}}]f,
\end{eqnarray}
where
\begin{eqnarray}
[X_{A,B},X_{\{E,C,D\},F}]f&=&
\{\{f,\{E,C,D\},F\},A,B\}-\{\{f,A,B\},\{E,C,D\},F\},\label{CL3}\\
\lbrack X_{A,B},X_{E,\{F,C,D\}}\rbrack f&=&
\{\{f,E,\{F,C,D\}\},A,B\}-\{\{f,A,B\},E,\{F,C,D\}\}.\label{CL4}
\end{eqnarray}
Here we should make two more choices 
on the use of the fundamental identity.
If we use the fundamental identities for the first terms in 
Eqs. (\ref{CL3}) and (\ref{CL4}), we get 
\begin{eqnarray}
&&\!\!\!\!\!\!\!\!\!\!\!\!\!\!\!\!\!\!\!\!\!\!\!\!\!\![X_{A,B},[X_{C,D},X_{E,F}]]f \nonumber\\
&&\!\!\!\!\!\!\!\!\!\!\!\!\!\!\!\!=\Big(
X_{\{\{E,C,D\},A,B\},F}+X_{\{E,C,D\},\{F,A,B\}}
+X_{\{E,A,B\},\{F,C,D\}}+X_{E,\{\{F,C,D\},A,B\}}
\Big)f.
\label{1AX}
\end{eqnarray}
There are a total of three two-choices for the use of the fundamental identity,
and therefore
the term $[X_{A,B},[X_{C,D},X_{E,F}]]f$ has $2^3=8$ types of expressions.
Expressions other than Eq. (\ref{1AX}) are as follows:
\begin{eqnarray}
&&\!\!\!\!\!\!\!\!\!\!\!\!\!\!\!\!\!\!\!\!\!\![X_{A,B},[X_{C,D},X_{E,F}]]f \nonumber\\
&&\!\!\!\!\!\!\!\!\!\!\!\!\!\!\!\!=\Big(
X_{\{\{E,C,D\},A,B\},F}+X_{\{E,C,D\},\{F,A,B\}}
-X_{\{A,E,\{F,C,D\}\},B}-X_{A,\{B,E,\{F,C,D\}\}}
\Big)f,
\label{1AY}\\
&&\!\!\!\!\!\!\!\!\!\!\!\!\!\!\!\!=\Big(
-X_{\{A,\{E,C,D\},F\},B}-X_{A,\{B,\{E,C,D\},F\}}
+X_{\{E,A,B\},\{F,C,D\}}+X_{E,\{\{F,C,D\},A,B\}}
\Big)f,
\label{1BX}\\
&&\!\!\!\!\!\!\!\!\!\!\!\!\!\!\!\!=\Big(
-X_{\{A,\{E,C,D\},F\},B}-X_{A,\{B,\{E,C,D\},F\}}
-X_{\{A,E,\{F,C,D\}\},B}-X_{A,\{B,E,\{F,C,D\}\}}
\Big)f,
\label{1BY}\\
&&\!\!\!\!\!\!\!\!\!\!\!\!\!\!\!\!=\Big(
-X_{\{\{C,E,F\},A,B\},D}-X_{\{C,E,F\},\{D,A,B\}}
-X_{\{C,A,B\},\{D,E,F\}}-X_{C,\{\{D,E,F\},A,B\}}
\Big)f,
\label{2AX}\\
&&\!\!\!\!\!\!\!\!\!\!\!\!\!\!\!\!=\Big(
-X_{\{\{C,E,F\},A,B\},D}-X_{\{C,E,F\},\{D,A,B\}}
+X_{\{A,C,\{D,E,F\}\},B}+X_{A,\{B,C,\{D,E,F\}\}}
\Big)f,
\label{2AY}\\
&&\!\!\!\!\!\!\!\!\!\!\!\!\!\!\!\!=\Big(
X_{\{A,\{C,E,F\},D\},B}+X_{A,\{B,\{C,E,F\},D\}}
-X_{\{C,A,B\},\{D,E,F\}}-X_{C,\{\{D,E,F\},A,B\}}
\Big)f,
\label{2BX}\\
&&\!\!\!\!\!\!\!\!\!\!\!\!\!\!\!\!=\Big(
X_{\{A,\{C,E,F\},D\},B}+X_{A,\{B,\{C,E,F\},D\}}
+X_{\{A,C,\{D,E,F\}\},B}+X_{A,\{B,C,\{D,E,F\}\}}
\Big)f.
\label{2BY}
\end{eqnarray}
Using some of the expressions given by Eqs. (\ref{1AX})-(\ref{2BY}), we can represent 
the second-order correction terms in $X_{\rm eff}$ (Eq. (\ref{BCH3}))
by a sum of Liouville operators. 
\par
If the shadow Hamiltonians exist, $X_{\rm eff}$ should be written as per Eq. (\ref{L2}).
In the case of the symplectic integrator for Hamiltonian mechanics, 
the effective Liouville operator $X_{\rm eff}$ can be written in the form of Eq. (\ref{L1})
and the shadow Hamiltonian $H_{S}$ is uniquely determined as in Eq. (\ref{SHPV}).
On the other hand, in the case of the structure-preserving integrator for Nambu mechanics,
it is nontrivial whether $X_{\rm eff}$ (Eq. (\ref{BCH3})) 
can be written in the form of Eq. (\ref{L2}).
It seems to depend on the model.
In the next section, we consider a simple model and try to find the shadow Hamiltonians.
\section{Example: $N=3$ harmonic oscillator}
Taking the $N=3$ harmonic oscillator presented in Sect. 3.2 as an example, 
let us construct the structure-preserving integrators
and derive the shadow Hamiltonians. 
\subsection{Structure-preserving integrators}
For the $N=3$ harmonic oscillator system,
two Hamiltonians (\ref{H})-(\ref{G}) are separable, 
\begin{eqnarray}
H(x_1,x_2,x_3)&=&A(x_2)+B(x_3),\\
G(x_1,x_2,x_3)&=&C(x_3)+D(x_1),
\end{eqnarray}
where 
\begin{eqnarray}
A(x_2)&=&\frac{1}{2m}x_2^{2},~~~~~B(x_3)~=~\frac{m\omega^2}{2}x_3,\label{AB}\\
C(x_3)&=&x_3,~~~~~~~~~~D(x_1)~=~-x_1^2.\label{CD}
\end{eqnarray}
Setting $H_1=0$, $H_2=A$, $H_3=B$, $G_1=D$, $G_2=0$, $G_3=C$,
the Liouville operator $X_{H,G}$ can be split into
$X_{H,G}=X_{1}+X_{2}+X_{3}$, where
\begin{eqnarray}
X_{1}&=&X_{A,C}~=~ \frac{1}{m}x_{2}\frac{\partial }{\partial x_{1}},\label{hX1}\\
X_{2}&=&X_{B,D}~=~ -m\omega^{2}x_{1}\frac{\partial }{\partial x_{2}},\label{hX2}\\
X_{3}&=&X_{A,D}~=~ \frac{2}{m}x_{1}x_{2}\frac{\partial }{\partial x_{3}}.\label{hX3}
\end{eqnarray}
Here $X_{1}$ and $X_{2}$ correspond to $X_{T}$ and $X_{V}$
in Eqs. (\ref{XTXV}), respectively.
Each Liouville operator gives the exact short time evolution as follows,
\begin{eqnarray}
e^{h X_{1}}\begin{pmatrix}x_{1}\\ x_{2}\\ x_{3}\end{pmatrix}\!\!&=&
\begin{pmatrix}
x_{1}+\frac{1}{m}h x_{2}\\ x_{2}\\ x_{3}
\end{pmatrix},\label{hX1e}\\
e^{h X_{2}}\begin{pmatrix}x_{1}\\ x_{2}\\ x_{3}\end{pmatrix}\!\!&=&
\begin{pmatrix}
x_{1}\\ x_{2}-m\omega^2 h x_{1}\\ x_{3}
\end{pmatrix},\label{hX2e}\\
e^{h X_{3}}\begin{pmatrix}x_{1}\\ x_{2}\\ x_{3}\end{pmatrix}\!\!&=&
\begin{pmatrix}
x_{1}\\ x_{2}\\ x_{3}+\frac{2}{m}h x_{1}x_{2}
\end{pmatrix}.\label{hX3e}
\end{eqnarray}
Composing five exact solutions symmetrically, 
we can construct a total of $3!=6$ types of 
second-order structure-preserving integrators:
$\Phi^{12321}_{h}$ (Eq. (\ref{12321})), $\Phi^{13231}_{h}$, $\Phi^{31213}_{h}$, 
$\Phi^{23132}_{h}$, $\Phi^{21312}_{h}$ and $\Phi^{32123}_{h}$.
It should be noted here that 
$X_{1}=X_{T}$ and $X_{2}=X_{V}$, and they do not depend on $x_3$.
That is, the time evolution of $x_3$ is decoupled from 
the time evolutions of $x_1$ and $x_2$, and 
the relationship with the symplectic integrator $\Phi^{TVT}_{h}$ or $\Phi^{VTV}_{h}$ 
for the harmonic oscillator can be written as follows:
\begin{eqnarray}
&&\Phi^{12321}_{h}\begin{pmatrix}x_{1}\\ x_{2}\end{pmatrix}=
\Phi^{13231}_{h}\begin{pmatrix}x_{1}\\ x_{2}\end{pmatrix}=
\Phi^{31213}_{h}\begin{pmatrix}x_{1}\\ x_{2}\end{pmatrix}=
\Phi^{TVT}_{h}\begin{pmatrix}x_{1}\\ x_{2}\end{pmatrix},\label{equiv1}\\
&&\Phi^{23132}_{h}\begin{pmatrix}x_{1}\\ x_{2}\end{pmatrix}=
\Phi^{21312}_{h}\begin{pmatrix}x_{1}\\ x_{2}\end{pmatrix}=
\Phi^{32123}_{h}\begin{pmatrix}x_{1}\\ x_{2}\end{pmatrix}=
\Phi^{VTV}_{h}\begin{pmatrix}x_{1}\\ x_{2}\end{pmatrix}.\label{equiv2}
\end{eqnarray}
\subsection{Shadow Hamiltonians}
Here we consider the integrator $\Phi^{12321}_{h}$.
Using $H_1=0$, $H_2=A$, $H_3=B$, $G_1=D$, $G_2=0$, $G_3=C$ and 
noting that some commutators are zero in the present case,
$[X_{1},[X_{1},X_{3}]]=[X_{2},[X_{2},X_{3}]]=[X_{3},[X_{3},X_{1}]]=[X_{3},[X_{3},X_{2}]]=
[X_{3},[X_{1},X_{2}]]=0$,
Eq. (\ref{BCH3}) can be simplified as
\begin{eqnarray}
X_{\rm eff}= X_{H,G}\!\!\!&-&\!\!\!\frac{h^2}{24}\Big(~~
[X_{A,C},[X_{A,C},X_{B,D}]]+[X_{A,C},[X_{B,D},X_{A,D}]]\nonumber\\
&&~-2[X_{B,D},[X_{B,D},X_{A,C}]]+[X_{B,D},[X_{A,C},X_{A,D}]]~~\Big)+O(h^{4}).\label{BCH4}
\end{eqnarray}
Then, by using the fundamental identity three times,
we have $2^3=8$ types of expressions of the double commutators
(\ref{1AX})-(\ref{2BY}).
For example, for the double commutator $[X_{A,C},[X_{A,C},X_{B,D}]]$ in Eq. (\ref{BCH4}), 
Eqs. (\ref{1AX}) and (\ref{1BX}) give $X_{B,-\frac{2}{m^2}x_2^2}$, 
whereas Eqs. (\ref{1AY}), (\ref{1BY})-(\ref{2BY}) give $X_{\frac{\omega^2}{m}x_2^2,C}$.
The other three double commutators in Eq. (\ref{BCH4})
also have two different representations.
The results can be summarized as follows:
\begin{eqnarray}
[X_{A,C},[X_{A,C},X_{B,D}]]&=&
X_{B,-\frac{2}{m^2}x_2^2} ~~~{\rm or}~~~ X_{\frac{\omega^2}{m}x_2^2,~\!C}~,\label{CL5}\\
\lbrack X_{A,C},[X_{B,D},X_{A,D}]\rbrack &=&
X_{A,~2\omega^2 x_1^2} ~~~~~\!\!{\rm or}~~~ X_{-\frac{\omega^2}{m}x_2^2,~\!D}~,
\label{CL6}\\
\lbrack X_{B,D},[X_{B,D},X_{A,C}]\rbrack &=&
X_{B,-2\omega^2 x_1^2} ~~~{\rm or}~~~ X_{m\omega^4 x_1^2,~\!C}~,\label{CL7}\\
\lbrack X_{B,D},[X_{A,C},X_{A,D}]\rbrack &=&
X_{A,~2\omega^2 x_1^2} ~~~~~\!\!{\rm or}~~~ X_{-\frac{\omega^2}{m}x_2^2,~\!D} ~.\label{CL8}
\end{eqnarray}
There are various possible representations of $X_{\rm eff}$.
If we choose 
\begin{eqnarray}
[X_{A,C},[X_{A,C},X_{B,D}]]&=&
3X_{B,-\frac{2}{m^2}x_2^2} -2X_{\frac{\omega^2}{m}x_2^2,~\!C}~,\label{CL5a}\\
\lbrack X_{A,C},[X_{B,D},X_{A,D}]\rbrack &=&X_{-\frac{\omega^2}{m}x_2^2,~\!D}~,
\label{CL6a}\\
\lbrack X_{B,D},[X_{B,D},X_{A,C}]\rbrack &=&X_{m\omega^4 x_1^2,~\!C}~,\label{CL7a}\\
\lbrack X_{B,D},[X_{A,C},X_{A,D}]\rbrack &=&X_{-\frac{\omega^2}{m}x_2^2,~\!D} ~,\label{CL8a}
\end{eqnarray}
then Eq. (\ref{BCH3}) can be rewritten as follows:
\begin{eqnarray}
X_{\rm eff}= X_{H,G}+X_{H,~\!\frac{1}{4m^2}h^2x_2^2}
+X_{\frac{\omega^2}{12m}h^2x_2^2+\frac{m\omega^4}{12}h^2 x_1^2,~\!G}
+O(h^{4}),\label{BCH5}
\end{eqnarray}
where we used $X_{A,-\frac{2}{m^2}x_2^2}=0$ and $X_{m\omega^4 x_1^2,~\!D}=0$.
Therefore,  if we choose (\ref{CL5a})-(\ref{CL8a})
the effective Liouville operator for the integrator $\Phi_{12321}(h)$
can be written as $X_{\rm eff}=X_{H_{S},G_{S}}$
with the shadow Hamiltonians:
\begin{eqnarray}
H_{S}&=& H+\frac{\omega^2}{12m}h^2x_2^2+\frac{m\omega^4}{12}h^2 x_1^2
+O(h^{4}),\label{SH12321a}\\
G_{S}&=& G+\!\frac{1}{4m^2}h^2x_2^2 
+O(h^{4}).\label{SG12321a}
\end{eqnarray}
Note that these shadow Hamiltonians are not unique.
For example, if we choose 
\begin{eqnarray}
[X_{A,C},[X_{A,C},X_{B,D}]]&=&
X_{B,-\frac{2}{m^2}x_2^2} ~,\label{CL5b}\\
\lbrack X_{A,C},[X_{B,D},X_{A,D}]\rbrack &=&
X_{A,~2\omega^2 x_1^2} ~,\label{CL6b}\\
\lbrack X_{B,D},[X_{B,D},X_{A,C}]\rbrack &=&
X_{B,-2\omega^2 x_1^2} ~,\label{CL7b}\\
\lbrack X_{B,D},[X_{A,C},X_{A,D}]\rbrack &=&
X_{A,~2\omega^2 x_1^2} ~,\label{CL8b}
\end{eqnarray}
and use $X_{A,-\frac{2}{m^2}x_2^2}=0$,
then we obtain other types of shadow Hamiltonians:
\begin{eqnarray}
H_{S}&=& H+O(h^{4}),\label{SH12321b}\\
G_{S}&=& G+\!\frac{1}{12m^2}h^2x_2^2 -\frac{\omega^2}{6}h^2x_1^2
+O(h^{4}).\label{SG12321b}
\end{eqnarray}
We refer to the shadow Hamiltonians calculated using the BCH formula 
as the BCH shadow Hamiltonians.
Although we could calculate the BCH Hamiltonians here, 
they have indefinite expressions.
\subsection{Conserved quantities and exact shadow Hamiltonians}
It can be shown by explicit calculation that the structure-preserving 
integrator $\Phi^{12321}_{h}$ conserves the following quantities,
\begin{eqnarray}
H_{c}&=&H,\label{CQ12321H}\\
G_{c}&=&G+\frac{1}{4m^2}h^2x_2^{2}.\label{CQ12321G}
\end{eqnarray}
Although $H_{c}$ and $G_{c}$ are conserved, they are not the shadow Hamiltonians.
That is, the Liouville operator associated with $H_c$ and $G_c$, 
\begin{eqnarray}
X_{H_c,G_c}= \frac{1}{m}\Big(1-\frac{\omega^2 h^2}{4}\Big)x_{2}
\frac{\partial }{\partial x_{1}}
-m\omega^{2}x_{1}\frac{\partial }{\partial x_{2}}
+\frac{2}{m}x_1x_2\frac{\partial}{\partial x_3},\label{XHcGc}
\end{eqnarray}
is not the effective Liouville operator $X_{\rm eff}$.
Since $\Phi^{12321}_{h}=\Phi^{TVT}_{h}$ for the time evolutions of 
$x_1$ and $x_2$ as shown in Eq. (\ref{equiv1}),
the $x$- and $y$-derivative terms in the effective Liouville operator 
$X_{\rm eff}$ for $\Phi^{12321}_{h}$ must agree with
the effective Liouville operator (\ref{Xeff1}) for $\Phi^{TVT}_{h}$.
From this consistency condition, we have
\begin{eqnarray}
X_{\rm eff}=F(\omega h)X_{H_{c},G_{c}}\label{Xeff2},
\end{eqnarray}
where the factor $F(x)$ is same as Eq. (\ref{factor1}).
The exact  shadow Hamiltonians $H_{S}^{e}$ and $G_{S}^{e}$ are given by
identifying $X_{\rm eff}=X_{H_{S}^{e},G_{S}^{e}}$.
However, 
depending on the way to distribute the factor $F(\omega h)$ to $H_{c}$ and $G_{c}$,
the exact shadow Hamiltonians take different forms.
In general, we can define them with  a real parameter $\alpha$,
\begin{eqnarray}
H_{S}^{e}\!&=&\!F(\omega h)^{\alpha}H_{c}
=H+\frac{\alpha\omega^2}{12m}h^2x_2^2+\frac{\alpha m\omega^4}{12}h^2 x_3
+O(h^{4}),\label{ESH12321}\\
G_{S}^{e}\!&=&\!F(\omega h)^{1-\alpha}G_{c}
=G+\frac{1}{4m^2}h^2x_2^2  + \frac{(1-\alpha)\omega^2}{6}h^2x_3 
- \frac{(1-\alpha)\omega^2}{6}h^2x_1^2
+O(h^{4}).\label{ESG12321}
\end{eqnarray}
On the other hand, for the BCH shadow Hamiltonians, 
we noted that
there is arbitrariness due to the choices for the use of the fundamental identity.
Here we represent the effective Liouville operator (\ref{BCH4}) as 
\begin{eqnarray}
X_{\rm eff}=\alpha X_{\rm eff}+(1-\alpha)X_{\rm eff} \label{Xeff3}
\end{eqnarray}
with a real parameter $\alpha$.
Using Eqs. (\ref{CL5a})-(\ref{CL8a}) for the first term $\alpha X_{\rm eff}$ and
using Eqs. (\ref{CL5b})-(\ref{CL8b}) for the second term $(1-\alpha)X_{\rm eff}$, 
the effective Liouville operator can be written as 
$X_{\rm eff}=X_{H_{S},G_{S}}$
with the shadow Hamiltonians:
\begin{eqnarray}
H_{S}\!&=&\!
H+\frac{\alpha\omega^2}{12m}h^2x_2^2+\frac{\alpha m\omega^4}{12}h^2 x_1^2
+O(h^{4}),\label{SH12321c}\\
G_{S}\!&=&\!
G+\frac{1}{12m^2}h^2x_2^2  + \frac{\alpha}{6m^2}h^2x_2^2
- \frac{(1-\alpha)\omega^2}{6}h^2x_1^2
+O(h^{4}).\label{SG12321c}
\end{eqnarray}
If we replace $x_3$ in the third term of Eq. (\ref{ESH12321})
using $G_{S}^{e}={\rm Const}$,
replace $x_3$ in the third term of Eq. (\ref{ESG12321})
using $H_{S}^{e}={\rm Const}$,
and neglect the constant terms,
then the BCH shadow Hamiltonians (\ref{SH12321c})-(\ref{SG12321c}) are consistent 
with the exact shadow Hamiltonians (\ref{ESH12321})-(\ref{ESG12321})  
up to the second order of $h$.
\par
Both of the BCH shadow Hamiltonians and the exact shadow Hamiltonians 
have indefinite forms parameterized by $\alpha$.
The parameter $\alpha$ is unphysical and 
vanishes in the effective Liouville operator $X_{\rm eff}$ (Eq. (\ref{Xeff2}) or (\ref{Xeff3})),
i.e.
the effective dynamics $df/dt=X_{\rm eff}f$ does not depend on the  parameter $\alpha$.
This indefiniteness would originate from the multi-Hamiltonian structure of the 
Nambu mechanics.
\subsection{Summary of the second-order integrators}
So far we have considered one of the second-order structure-preserving integrators
$\Phi^{12321}_{h}$ for the $N=3$ harmonic oscillator 
described by the three variables  (\ref{Ex1}).
We have calculated the BCH shadow Hamiltonians, and found conserved quantities and the exact shadow Hamiltonians.
Similar analyses can be performed for the other five integrators.
In Table \ref{table} we summarize the results of 
the conserved quantities $H_{c}$ and $G_{c}$ for six integrators. 
Note that the conserved quantities are the same 
in $\Phi^{12321}_{h}$ and $\Phi^{13231}_{h}$.
This is because they give the same time evolution,
$\Phi^{12321}_{h}\bm{x}(t)=\Phi^{13231}_{h}\bm{x}(t)$.
For the same reason, the conserved quantities in the integrator $\Phi^{21312}_{h}$ 
are same as those in $\Phi^{23132}_{h}$.
The exact shadow Hamiltonians have indefinite expressions, 
and they can be given by $H_{S}^{e}=F(\omega h)^{\alpha}H_{c}$ 
and $G_{S}^{e}=F(\omega h)^{1-\alpha}G_{c}$ with a real parameter $\alpha$.

\begin{table}[!h]
\caption{Conserved quantities $H_{c}$ and $G_{c}$ of the second-order integrators
for the $N=3$ harmonic oscillator
described by variables $(x_1, x_2 , x_3)=(q, p, q^2)$.
The exact shadow Hamiltonians can be given by $H_{S}^{e}=F(\omega h)^{\alpha}H_{c}$ 
and $G_{S}^{e}=F(\omega h)^{1-\alpha}G_{c}$ with a real parameter $\alpha$.}
\label{table}
\centering
{\renewcommand\arraystretch{2.0}
\tabcolsep = 1cm
\begin{tabular}{|c||c|c|}
\hline
 & $H_{c}$ & $G_{c}$\\ 
\hline
$\Phi^{12321}_{h}$ & $H$ & $G+\frac{1}{4m^2}h^2x_2^{2}$\\
\hline
$\Phi^{13231}_{h}$ & $H$ & $G+\frac{1}{4m^2}h^2x_2^{2}$\\
\hline
$\Phi^{31213}_{h}$ & $H-\frac{\omega^2}{4m}h^2x_2^{2}$ & 
$G-\frac{1}{4m^2}h^2x_2^{2}$\\
\hline
$\Phi^{21312}_{h}$ & $H-\frac{m \omega^4}{8}h^2x_1^{2}$ & $G$\\
\hline
$\Phi^{23132}_{h}$ & $H-\frac{m \omega^4}{8}h^2x_1^{2}$ & $G$\\
\hline
$\Phi^{32123}_{h}$ & $H+\frac{m \omega^4}{8}h^2x_1^{2}$ &
$G+\frac{\omega^2}{2}h^2x_1^{2}$\\
\hline
\end{tabular}
}
\end{table}
\subsection{Numerical results}
\begin{figure}[htbp]
  \begin{minipage}[b]{0.45\linewidth}
    \centering
    \includegraphics[bb=0 0 640 384, scale=0.33]{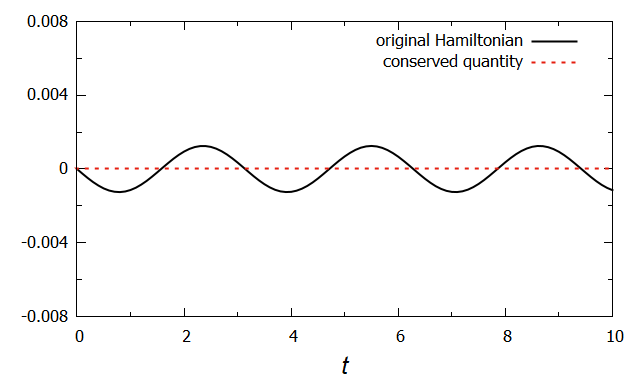}
    \caption{Time evolutions by $\Phi^{32123}_{h}$:
The error in the original Hamiltonian $H(t)-H(0)$ (solid line) and 
the error in the conserved quantity $H_{c}(t)-H_{c}(0)$ (dashed line).}
\label{Fig1}
  \end{minipage}
\hspace{8mm}
  \begin{minipage}[b]{0.45\linewidth}
    \centering
    \includegraphics[bb=0 0 640 384, scale=0.33]{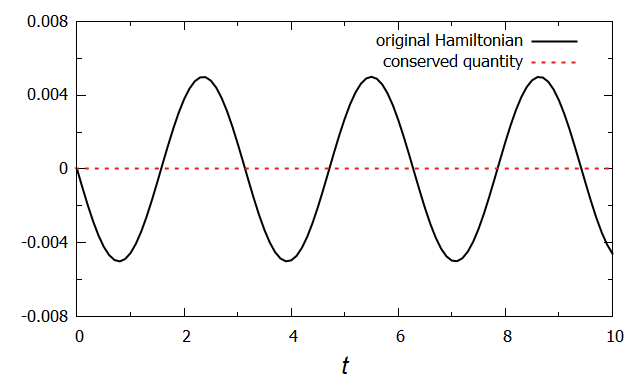}
    \caption{Time evolutions by $\Phi^{32123}_{h}$:
The error in the original Hamiltonian $G(t)-G(0)$ (solid line) and 
the error in the conserved quantity $G_{c}(t)-G_{c}(0)$ (dashed line).}
\label{Fig2}
  \end{minipage}
\end{figure}

Finally we give a numerical demonstration of a second-order integrator
constructed in this work.
We present here the results of one of the six integrators, $\Phi^{32123}_{h}$.
The original Hamiltonians $H$ and $G$ are given in Eqs. (\ref{H})-(\ref{G}),
the conserved quantities are 
$H_{c}=H+\frac{m \omega^4}{8}h^2x_1^{2}$ and 
$G_{c}=G+\frac{\omega^2}{2}h^2x_1^{2}$,
and the exact shadow Hamiltonians can be written as 
 $H_{S}^{e}=F(\omega h)^{\alpha}H_{c}$ 
and $G_{S}^{e}=F(\omega h)^{1-\alpha}G_{c}$ with a real parameter $\alpha$.
Here we set the parameters $m=\omega=1$ and the small time step $h=0.1$.
We evolve the harmonic oscillator in time by $n$ steps using the integrator 
$\Phi^{32123}_{h}$ with initial conditions $(x_{1}(0),x_{2}(0),x_{3}(0))=(1,1,1)$,
and obtain numerical solutions $\bm{x}^{32123}(t)=(\Phi^{32123}_{h})^{n}\bm{x}(0)$,
where $t=nh$.
Then using the solutions $\bm{x}^{32123}(t)$ we evaluate the original Hamiltonians $H(t)$ and $G(t)$ and the conserved quantities $H_{c}(t)$ and $G_{c}(t)$.
Figure \ref{Fig1} shows the error in the original Hamiltonian, $H(t)-H(0)$, and 
the error in the conserved quantity, $H_{c}(t)-H_{c}(0)$.
Figure \ref{Fig2} also shows the errors in the original Hamiltonian, $G(t)-G(0)$, and
in the conserved quantity, $G_{c}(t)-G_{c}(0)$.
It can be seen that neither $H(t)$ nor $G(t)$ is conserved; however, 
either of them oscillates with no increase in error.
That is, the integrator $\Phi^{32123}_{h}$ gives stable time evolution and 
can be used for long-time simulations as can symplectic integrators 
for Hamiltonian mechanics.
\par  
\par\vspace{0mm}
It can also be seen from  Figs. \ref{Fig1}-\ref{Fig2} 
that $H_{c}$ and $G_{c}$ are both conserved. 
However, note that they are not the shadow Hamiltonians,
but just conserved quantities.
To see this point clearly, 
we calculate the exact solutions of the Nambu equations with 
original Hamiltonians, the conserved quantities, and the exact shadow Hamiltonians,
and compare those three solutions 
with the numerical solutions computed by the integrator $\Phi^{32123}_{h}$.
The Nambu equations with the original Hamiltonians  $H$ and $G$ are given by 
Eqs. (\ref{NambuOHeqs}).
On the other hand, the Nambu equations with the conserved quantities  
$H_{c}$ and $G_{c}$ as the Hamiltonians are written as
\begin{eqnarray}
\frac{d}{dt}x_{1}=\frac{1}{m}x_{2},~~~~~~
\frac{d}{dt}x_{2}=-m\omega^{2}a^{2}x_{1},~~~~~~
\frac{d}{dt}x_{3}=\frac{2}{m}b^{2}x_{1}x_{2},
\label{NambuCQeqs}
\end{eqnarray}
where $a=\sqrt{1-\frac{\omega^2 h^2}{4}}$ and $b=\sqrt{1-\frac{\omega^2 h^2}{2}}$.
Then, the Nambu equations 
with the exact shadow Hamiltonians  $H_{S}^{e}$ and $G_{S}^{e}$ are given by
\begin{eqnarray}
\frac{d}{dt}x_{1}=\frac{1}{m}F(\omega h)x_{2},~~~~
\frac{d}{dt}x_{2}=-m\omega^{2}a^{2}F(\omega h)x_{1},~~~~
\frac{d}{dt}x_{3}=\frac{2}{m}b^{2}F(\omega h)x_{1}x_{2}.
\label{NambuSHeqs}
\end{eqnarray}
\par
\begin{figure}[htbp]
  \begin{minipage}[b]{0.45\linewidth}
    \centering
    \includegraphics[bb=0 0 640 384, scale=0.33]{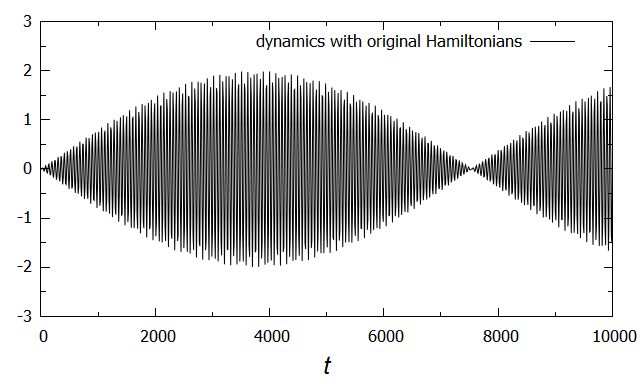}
    \caption{The difference between the two dynamics: 
$x^{\rm o}_{3}(t)-x^{32123}_{3}(t)$ (solid line). Here data are plotted every 100 steps.}
\label{Fig3}
  \end{minipage}
\hspace{8mm}
  \begin{minipage}[b]{0.45\linewidth}
    \centering
    \includegraphics[bb=0 0 640 384, scale=0.33]{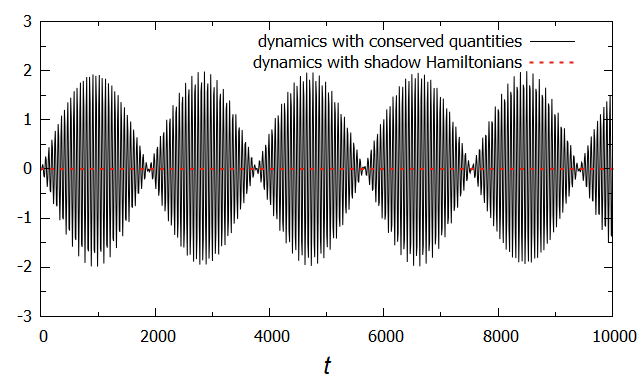}
    \caption{The differences between the two dynamics: 
$x^{\rm c}_{3}(t)-x^{32123}_{3}(t)$ (solid line) and 
$x^{\rm s}_{3}(t)-x^{32123}_{3}(t)$ (dashed line).}
\label{Fig4}
  \end{minipage}
\end{figure}
\par
We denote the solutions of Eqs. (\ref{NambuOHeqs}) by $\bm{x}^{\rm o}(t)$,
the solutions of Eqs. (\ref{NambuCQeqs}) by $\bm{x}^{\rm c}(t)$, and
the solutions of Eqs. (\ref{NambuSHeqs}) by $\bm{x}^{\rm s}(t)$.
Here we give only the third components of the solutions,
\begin{eqnarray}
x_{3}^{\rm o}(t)&=&\Big(\frac{1}{2}x_{1}(0)^2-\frac{1}{2m^2\omega^2}x_{2}(0)^2\Big)
\cos2\omega t 
+\frac{1}{m\omega}x_{1}(0)x_{2}(0)\sin2\omega t \nonumber\\
&&~+x_{3}(0)-\frac{1}{2}x_{1}(0)^2+\frac{1}{2m^2\omega^2}x_2(0)^{2},\label{OHsol}\\
x_{3}^{\rm c}(t)&=&b^{2}\Big[\Big(\frac{1}{2}x_{1}(0)^2-\frac{1}{2m^2\tilde{\omega}^2}
x_{2}(0)^2\Big)\cos2\tilde{\omega} t 
+\frac{1}{m\tilde{\omega}}x_{1}(0)x_{2}(0)\sin2\tilde{\omega} t \nonumber\\
&&~~~~~+\frac{1}{b^2}x_{3}(0)-\frac{1}{2}x_{1}(0)^2+\frac{1}{2m^2\tilde{\omega}^2}x_2(0)^{2} \Big],
\label{CQsol}\\
x_{3}^{\rm s}(t)&=&b^{2}\Big[\Big(\frac{1}{2}x_{1}(0)^2-\frac{1}{2m^2\tilde{\omega}^2}
x_{2}(0)^2\Big)\cos2\tilde{\omega} \tilde{t} 
+\frac{1}{m\tilde{\omega}}x_{1}(0)x_{2}(0)\sin2\tilde{\omega} \tilde{t} \nonumber\\
&&~~~~~+\frac{1}{b^2}x_{3}(0)-\frac{1}{2}x_{1}(0)^2+\frac{1}{2m^2\tilde{\omega}^2}x_2(0)^{2} \Big],
\label{SHsol}
\label{Solutions}
\end{eqnarray}
where $\tilde{\omega}=a\omega$ and $\tilde{t}=F(\omega h)t$.
Figure \ref{Fig3} shows the difference between the exact solution of
the Nambu mechanics with the original Hamiltonians (\ref{OHsol}) and
the numerical solution given by the integrator $\Phi^{32123}_{h}$, 
$x^{\rm o}_{3}(t)-x^{32123}_{3}(t)$. 
Although both of $x^{\rm o}_{3}(t)$ and $x^{32123}_{3}(t)$ are periodic, 
the frequencies are slightly different and we can 
observe a beat over a very long time domain.
Figure \ref{Fig4} show two more differences, $x^{\rm c}_{3}(t)-x^{32123}_{3}(t)$ and 
$x^{\rm s}_{3}(t)-x^{32123}_{3}(t)$.
Since a beat is also observed in the difference $x^{\rm c}_{3}(t)-x^{32123}_{3}(t)$,
the conserved quantities $H_{c}$ and $G_{c}$ are not the shadow Hamiltonians 
to reproduce the time evolution given by the integrator $\Phi^{32123}_{h}$.
On the other hand, it can be seen that $x^{\rm s}_{3}(t)-x^{32123}_{3}(t)$ remains zero
all the time, indicating that $H_{S}^{e}$ and $G_{S}^{e}$ work as the shadow Hamiltonians
of this integrator.

\section{Conclusions and future works}
\label{Conclusions}
We have presented a general procedure to construct the structure-preserving integrators 
for Nambu mechanics and calculate the shadow Hamiltonians. 
Taking the $N=3$ harmonic oscillator as an example, 
we have shown that the shadow Hamiltonians 
can be calculated by using the BCH formula and the fundamental identity,
and also the exact expressions of the shadow Hamiltonians can be found
in terms of the consistency with the exact shadow Hamiltonian of the symplectic integrator.
This is the first work to derive the explicit forms of BCH shadow Hamiltonians 
and exact shadow Hamiltonians in a Nambu system.
However, we have also found that they both have indefinite expressions.
The BCH shadow Hamiltonians are indefinite 
due to the freedom in the choices for the use of the fundamental identity,
whereas the exact shadow Hamiltonians are indefinite 
due to the way of distribution of the factor $F(\omega h)$.
Although they are both indefinite, 
we have confirmed that they are consistent with each other up to the order $h^{2}$.
\par
In the present paper, we have focused on second-order structure-preserving integrators 
and given a simple example where the shadow Hamiltonians exist.
However, it has not been clarified what kind of Nambu systems and integrators would 
have shadow Hamiltonians.
It would be interesting to study other Nambu systems
and other types of integrators.
\par
In the Hamiltonian mechanics,
it has been an important problem to construct symplectic integrators 
in nonseparable Hamiltonian systems (see, e.g. Ref. \cite{Tao}).
In the Nambu mechanics, we often see nonseparable 
Hamiltonian(s) in multiple Hamiltonians. For example, 
the $N=3$ harmonic oscillator with three composite variables, 
$(x_1, x_2 , x_3)=(q^{2}, p^{2}, qp)$, is driven by the Hamiltonians
$H=\frac{1}{2m}x_2+\frac{m\omega^2}{2}x_1$ and $G=2x_3^2-2x_1x_2$ \cite{Horikoshi1}.
And the cyclic Lotka-Volterra model with three species $(x_1, x_2 , x_3)$ is driven 
by the Hamiltonians $H=x_1+x_2+x_3$ and $G=x_{1}x_{2}x_{3}$ 
\cite{FrachebourgKrapivskyBenNaim}.
In both systems, the Hamiltonian $G$ is nonseparable.
It would be a challenging work to construct 
structure-preserving integrators for Nambu systems with nonseparable 
Hamiltonian(s).
\par
We have seen that the fundamental identity plays an important role 
in calculating the shadow Hamiltonians using the BCH formula.
However, it is well known that the fundamental identity does not hold
in many degrees of freedom systems \cite{Takhtajan,HoMatsuo,Horikoshi2}.
Therefore it would be interesting to see if shadow Hamiltonians 
exist in many degrees of freedom Nambu systems.
\par
The integrators we have constructed in the present paper are those  
with time-reversal symmetry.
It would be important to consider the role of time-reversal symmetry 
in structure-preserving integrators for the Nambu mechanics \cite{Modin}.
\par
Finally, it is known that the Nambu mechanics can always be represented 
in the form of the noncanonical Hamiltonian mechanics with the 
noncanonical Poisson bracket defined by the Nambu bracket \cite{Takhtajan,Horikoshi2,BialynickiBirulaMorrison}.
Therefore, it would also be interesting to study the relationship between 
the structure-preserving integrators for Nambu mechanics and 
the Lie-Poisson integrators for noncanonical Hamiltonian mechanics 
\cite{Karasozen,HairerLubichWanner,McLachlan}. 
\section*{Acknowledgments}
We thank the referee for letting us know about the cyclic Lotka-Volterra model.
This work was partly supported by a scientific grant 
from the Ministry of Education, Culture,
Sports, Science and Technology (MEXT) under Grant No.~26610134.


\end{document}